# Full-Spectrum Flexible Color Printing at the Diffraction Limit

*Patrizia Richner, Patrick Galliker, Tobias Lendenmann, Stephan J. P. Kress, David K. Kim, David J. Norris, Dimos Poulikakos*

P. Richner, P. Galliker, T. Lendenmann, D. Poulikakos

Laboratory for Thermodynamics in Emerging Technologies, ETH Zurich, Sonneggstrasse 3, 8092 Zurich, Switzerland

Email: dpoulikakos@ethz.ch

S.J.P. Kress, D.K. Kim, D.J. Norris

Optical Materials Engineering Laboratory, ETH Zurich, Leonhardstrasse 21, 8092 Zurich, Switzerland

*Abstract:* Color printing at the diffraction limit has been recently explored by fabricating nanoscale plasmonic structures with electron beam lithography. However, only a limited color range and constant intensity throughout the structure have been demonstrated. Here we show an alternative, facile approach relying on the direct, open-atmosphere electrohydrodynamic rapid nanodrip printing of controlled amounts of red, green and blue (RGB) quantum dots at a resolution of 250 nm. The narrow emission spectrum of the dots allows the coverage of a very broad color space, exceeding standard RGB (sRGB) of modern display devices. We print color gradients of variable intensity, which to date could not be achieved with diffraction-limited resolution. Showcasing the capabilities of the technology, we present a photo-realistic printed image of a colorful parrot with a pixel size of 250 nm.

Colloidal quantum dots have size-dependent emission spectra.[1] Since they can be stably dispersed in solution, they are ideal candidates for ink-jet and microcontact printing, as has been shown for a palette of applications ranging from light-emitting devices,[2-4] to displays[5] and photodetectors[6]. However, the resolution of these technologies is limited to a few microns. The diffraction limit of approximately 250 nm represents the smallest resolution at which color printing is resolvable with optical microscopes and therefore constitutes the lowest resolution limit for all related applications, such as anti-counterfeiting and optical data storage. Several groups have shown diffraction-limited color printing not based on dyes or quantum dots but by means of plasmonic structures.[7-12] They employed metal-insulator-metal (MIM) structures, which selectively reflect a portion of the incident light, depending on the structure size and shape. The drawbacks of this approach are on the one hand the limited color range compared to modern displays[9] and on the other hand the electron beam lithography steps necessary for the fabrication of the nanometer-sized structures, rendering the process highly expensive and difficult to upscale. Furthermore, for photorealistic color printing, it is necessary not only to generate a large range of hues but also to deliver a variation of intensity of those hues from dark to bright.

We have recently introduced an electrohydrodynamic printing technology termed NanoDrip printing,[13] capable of printing three-dimensional structures with a critical dimension of less than 100 nm, and placement precision down to a few nanometers[14] in an open atmosphere. The deposited materials range from metal nanoparticles[15] and metal salts[16] to a variety of quantum dots.[14, 17-19] Printing on a number of rigid[14, 17] and flexible substrates[19] has been shown.

In this work we present a facile method to generate full-color printed images with a diffraction-limited resolution based on red, green and blue colloidal quantum dots, representing an RGB color space analogous to the one applied in modern displays, with a color gamut far exceeding the standard RGB (sRGB) norm. By employing the NanoDrip technique,[13, 17, 20] we overcome limitations set by previous mask-based, plasmonic techniques by massively increasing the range of colors while maintaining high flexibility and low cost, which opens up promising pathway for cost-effective upscaling. The method employed here presents an approach to diffraction-limited color printing completely different from earlier works, as it does not rely on any plasmonic effects but combines the high deposition precision with the emission properties of quantum dots.

Figure 1 shows a schematic of the printing setup used in this work, described in detail in the methods section. As the ejection frequency of droplets from the nozzle remains constant, the amount of deposited material is solely defined by the speed at which the substrate is moved. In this manner, the amount of material deposited at a certain location can be precisely controlled, since more droplets are deposited per area at a lower stage speed and less at higher stage speeds. The intensity of the emitted light by colloidal quantum dots covering a certain area is a function of the number of quantum dots deposited in that area. By varying the speed of the stage the emitted light intensity can hence be controlled.

The three different quantum dot dispersions used here are all synthesized in-house; the synthesis procedures as well as size and other properties for the red, green and blue dots are described in previous publications.[14, 18-19] We use CdSe-CdS-ZnS core-shell-shell semiconductor nanocrystals for the red and green quantum dots, the blue quantum dots are CdS-ZnS core-shell nanocrystals with a composition gradient. The quantum dots feature emission peaks at 461 nm, 538 nm and 625 nm for a blue, green and red color, respectively. Detailed absorption and emission spectra can be found in Figures 2 a, b. The dilution of the dispersions in tetradecane is adjusted to ensure that very few quantum dots are contained in one ejected droplet, enabling fine tuning of the amount of deposited quantum dots and, consequently, of the emission intensity.

Figures 2c-e show photographs of printed, monochrome gradients for red, green and blue. Continuous lines are printed, varying the speed from slow on the left to very fast on the right, with a line-to-line spacing of 250 nm. The printing speed is defined for every individual 250 nm long section of a printed line, thus creating a pixel size of 250 nm. Arbitrary monochrome patterns with a diffraction-limited resolution can easily be printed. Since the speed of the stage could in theory be varied in incrementally small steps (~10 nm), we chose to employ 256 speed levels in analogy to the available brightness levels of an 8-bit image, where the brightness can be incremented from 0 to 255. The ability to control the brightness of every color from their saturated value to black sets our method apart from color printing techniques based on plasmonic MIM structures, where to our knowledge no variation in color intensity has been shown.

Figure 3a shows the CIE 1931 chromaticity diagram[21]. The colored area encompasses the color range perceivable by the human eye. The curved edge represents single-wavelength spectral colors, moving inwards, the colors are generated by a more broadband spectrum. The white, dashed triangle marks the amount of colors that can be displayed within the sRGB gamut of standard display technology.[22] It is limited by the broad emission spectrum of the RGB pixels in a display device. The quantum dots have a much narrower emission spectrum, hence their pure emission is located closer to the edge of the CIE 1931 gamut and spans a considerably larger color gamut than sRGB, in particular in the area of green wavelengths, where the human eye is most sensitive to color changes. This enables the generation of more hues.

The edge of the color gamut is represented by mixing the two colors of the adjacent corners. For each color combination (green-blue, red-blue and red-green) a gradient was printed on a glass substrate by subsequently the two different needed inks sequentially. The spectra of the three corresponding color gradients are displayed in Figures 3b-d. The spectral intensity over the wavelength is plotted along with the printed gradient in the y direction. A photograph of the printed gradient is shown to the left of each spectral graph, an arrow marks the y direction. While, for example, the apparent color in the middle of the red-green gradient is yellow, the spectrum shows that it is actually a mix of red and green with varying intensity (Figure 3d). Since the photographs are displayed in the conventional sRGB color space, not all the colors of the gradients shown here can be represented in their true hue, because they lie outside of the sRGB gamut.

The full capability of the color printing simultaneously at the diffraction limit and with a wide color gamut is best exemplified with a colorful microscale picture. The parrot shown in Figure 4a has a good representation of different colors, ranging from the pure red, blue and green to mixes such as yellow and teal as well as bright white, represented by an even mix of all three base colors. The printed image has a line-to-line spacing of 250 nm, the speed at which the stage is moved under the printing nozzle is defined for every 250 nm segment, leading to distinctive pixels with a pixel-to-pixel distance of 250 nm. The overall height of the image is 94 x 125 µm, close to the size of the cross section of a human hair. Figure 4b shows the image after the first printing step, displaying how the intensity of the red color is tuned. The green and blue quantum dots are printed sequentially on top of the red dots, following an alignment procedure, details of which are given in the Supplementary Information (Supporting Figure 1). The photograph in Figure 4c shows the final full-color printed parrot. However, these photographs can be merely viewed as a representation of the true richness of the real colors since the gamut of the printed image exceeds the gamut of the camera, in particular for green hues.

Figure 4d shows a scanning electron microscope image of the parrot's head after all three colors have been printed. The eye and the fine features of the feathers around the eye as well as the top-down orientation of the printed lines can be distinguished. Figure 4e shows a high magnification image of the area indicated in Figure 4d. It shows the transition from the dark background to the bright coat of the bird. The fast deceleration of the stage within a few hundred nanometers enables fasttransitions between fast and slow segments, leading to sharp controasts. When the stage moves slowly the spacing between the lines is not perfectly regular, as can be seen in Figure 4e. This effect is induced by the distortion of the electric field by previously printed lines. It could be avoided by depositing less material but this would result in a lower brightness of the printed pictures.

In summary, we report a relatively simple, direct method based on EHD NanoDrip printing, to generate full color images with a diffraction limited resolution based on red, green and blue colloidal quantum dots and a color gamut far exceeding the range of the sRGB norm used for displays. The capability to control the amount of deposited material allows the arbitrary mixing of colors to render different hues and simultaneously enables precise, pixel-wise brightness control.

We envision future implementations of the technology described here in applications targeted primarily at the nanoscale. Examples include transparent electrodes for touch screen devices, in- and out-of-plane plasmonic metal nanostructures and quantum dot emitter coupling with nanostructures. Further, targeting full device fabrication using solely additive manufacturing, going across scales from macro- to micro- and to nanoscale features, is a strong driver toward integration of NanoDrip in 3D printing platforms, to fill the gap from micro to nano feature resolution. We see also potential of integration of the NanoDrip process in multiscale surface mounting platforms, for example by implementing color-printed security features for high-end products as an anti-counterfeiting measure. In order to bridge the gap between the research-grade printing setup and its implementation in industry, upscaling as well as the integration of nanoscale alignment tools into macroscale production lines need to be addressed.

*Methods*

*Inks*: The inks are synthesized in-house, following published protocols.[14, 19] All colors are dispersed in n-tetradecane and are concentration-adjusted for an optical density of 5.0 at the first exciton. They are further diluted to a ratio of 1:10 (red), 1:7 (green) and 1:6 (blue) for printing. This accounts for the fact that the brightness of the red dots (undiluted) is higher than the green and blue ones. For the RGB image, the brightness of all colors has to be homogeneous. By adjusting the dilution we could guarantee that the brightness of the smallest amount of deposited quantum dots, i.e. the contents of one droplet corresponds to the same brightness irrespective of the emitted color.

*Printing*: Briefly, a glass capillary is pulled into the shape of a nozzle with an opening diameter of 1000-1100 nm with a Sutter Instruments pipette puller, which is coated with 10 nm Ti and 100 nm Au to render it conductive. The capillary is filled with the quantum-dot dispersion and brought within 5 µm of the glass substrate, which lies on a grounded holder, mounted on a piezo stage (not shown in the schematic). The solution is driven to the front of the nozzle by wetting. By applying a dc voltage of 225 V, droplets with a diameter down to ten times smaller than the nozzle opening are ejected regularly and rapidly from the apex of a larger stable meniscus hanging at the opening of the nozzle at a frequency of about 100 Hz. The detailed working principle can be found elsewhere.[13, 18] The substrate is then moved by the piezo-stage under the nozzle at a speed controllable by an in-house built control unit. The minimum printing speed was 1.3 µm/s, the maximum 141 µm/s. This ensures a maximal contrast between dark and bright areas while at the same time accounting for the limited acceleration capabilities of the piezo-stage. If the maximal speed were selected to be too high, sharp changes from bright to dark pixels would be blurred due to the finite acceleration. The details of the conversion from the RGB value to the speed are given in the Supplementary Information.

*Microscopy and Spectroscopy*: All photographs were taken with a Nikon 600 commercial camera mounted on a Nikon Eclipse LV100 microscope. The emission spectra were taken with an Andor Shamrock 303i spectrometer through a 200 µm slit at a sensor temperature of -75°C.

*Supporting Information Available:* Color alignment scheme and RGB value to stage speed conversion (PDF). The Supporting Information is available free of charge on the ACS Publications website at DOI: 10.1021/acsphotonics.XXXXXXX.

Notes

The authors declare the following competing financial interest(s): Two of the authors (P.G.,D.P.) are involved in a startup company that is attempting to commercialize the printing process used in our manuscript.

Acknowledgements

The research leading to these results has received funding from the Swiss National Science Foundation under Grant 146180. We also gratefully acknowledge funding from the European Research Council under the European14 Union's Seventh Framework Programme (FP/2007-2013) / ERC Grant Agreement Nr. 339905 (QuaDoPS Advanced Grant).

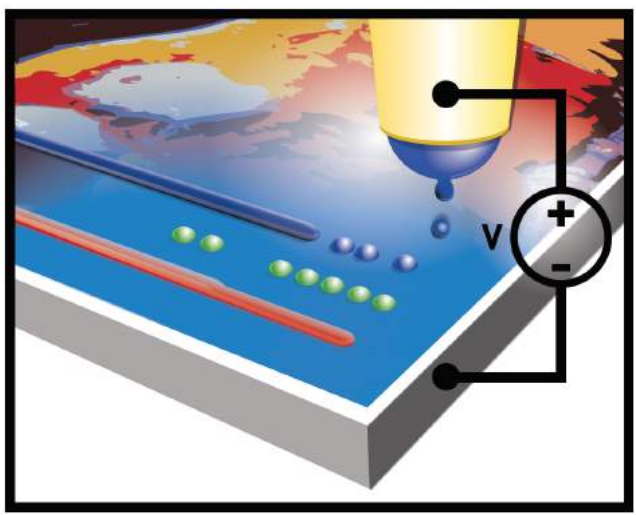

**Figure 1**: Schematic of the printing setup used in this work. The colloidal quantum-dot dispersion is printed with rapidly generated nanodrops from the apex of a larger stable meniscus at the nozzle, at 225 V. A controllable amount of quantum dots is deposited with high precision, this amount being a function of the substrate speed under the stationary nozzle.

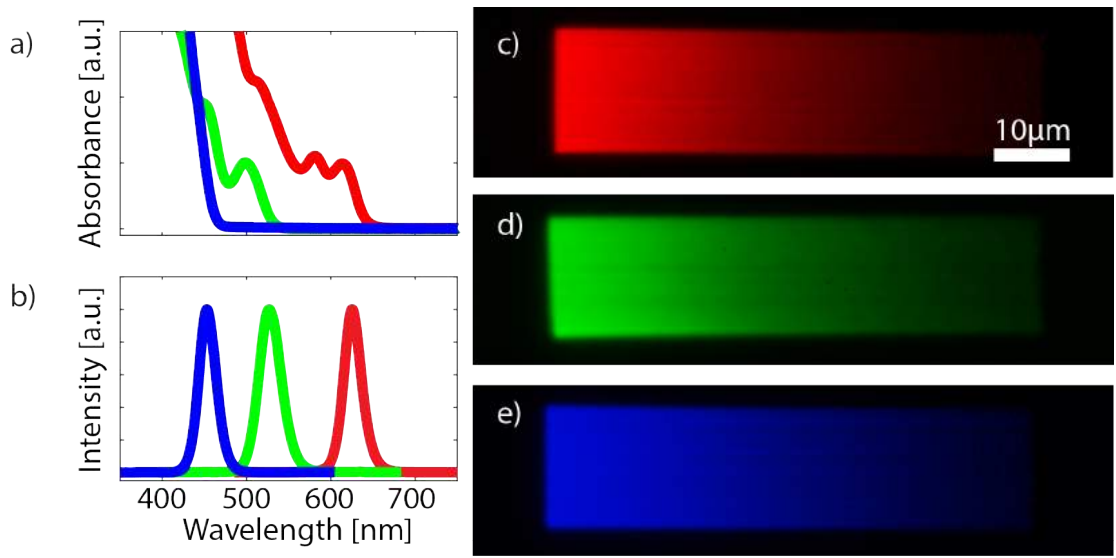

**Figure 2**: a) Absorption and b) emission spectra of the printed quantum dots; c)-e) Photographs of monochromatic gradients with varying intensity. The intensity is a function of the amount of deposited material, which in turn depends on the speed of the substrate relative to the nozzle during printing. The horizontal line-to-line spacing is 250 nm.

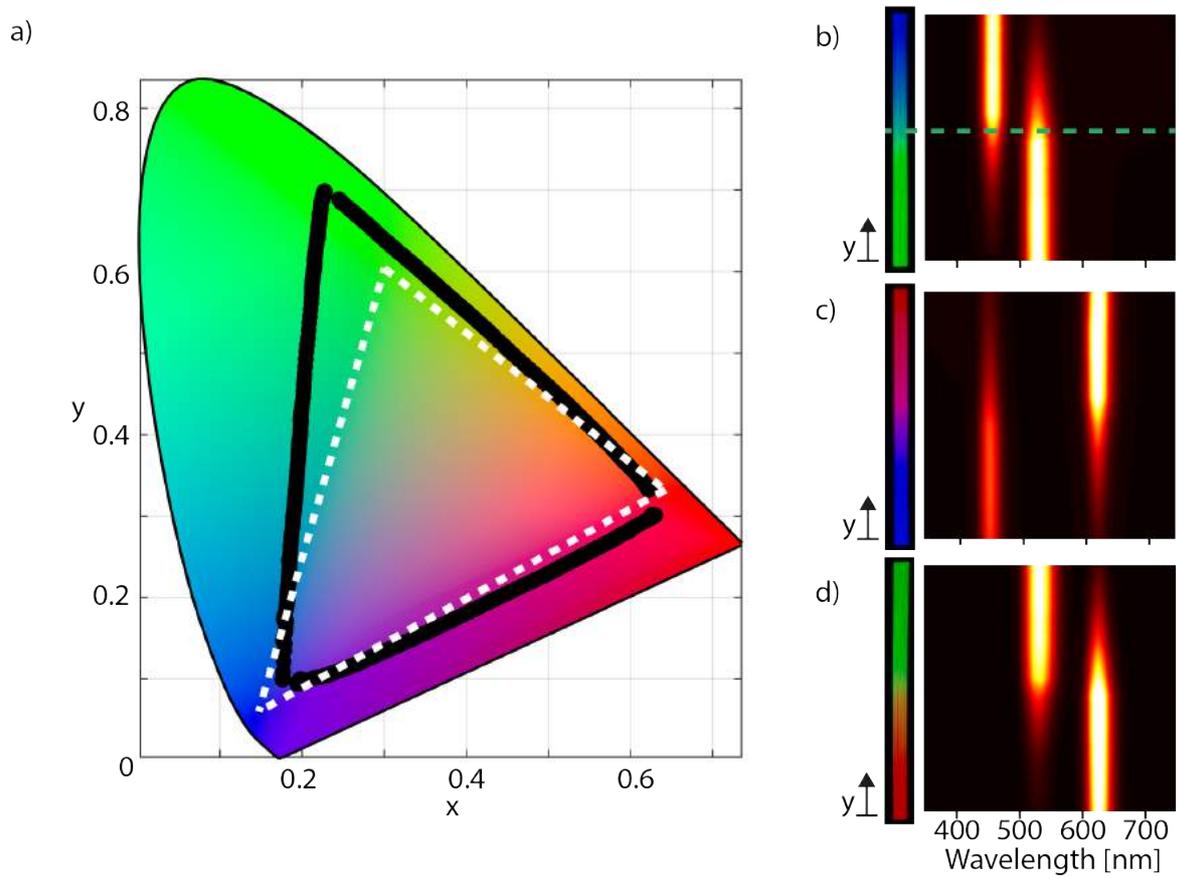

**Figure 3**: a) CIE chromaticity diagram: the white dashed line shows the sRGB colors that can be generated by conventional display devices. The black lines are calculated from the spectroscopy measurements of the two-color gradients shown in b)-d). A photograph of the printed gradients (7.5 x 130 µm) is shown next to the spectral graphs. The horizontal dashed line in b) is a guide showing of the correspondence of printed color and spectral intensity.

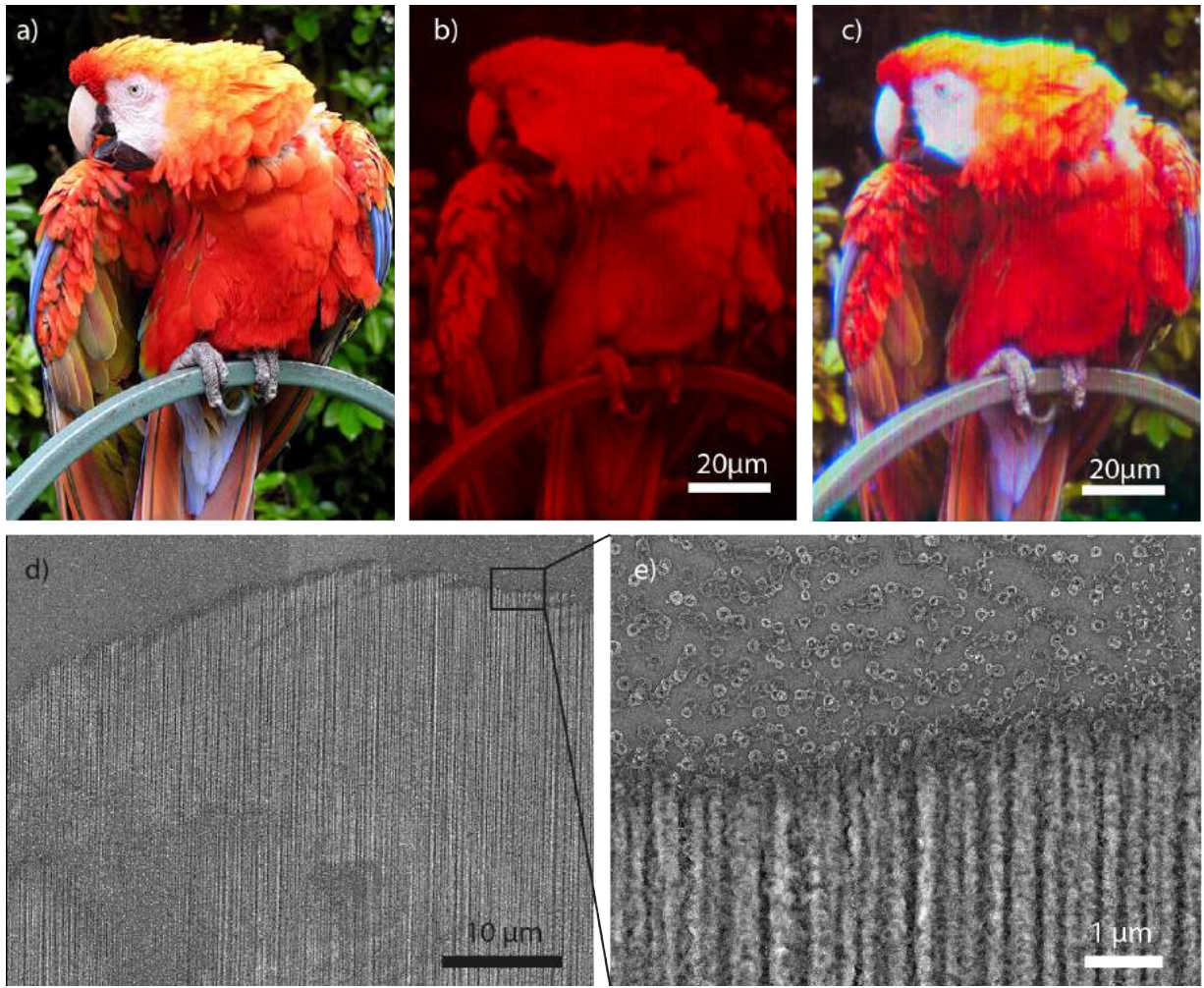

**Figure 4**: a) photograph of a colorful parrot; b) deposition of one color for an RGB image; c) full-color RGB image of a printed parrot, the total image size is 94x125 µm, roughly the size of the cross section of a human hair; d,e) SEM micrographs of a parrot's head printed with red, green and blue quantum dots. d) the eye as well as the structure of the feathers on the head can be distinguished and the orientation of the printed lines which form the picture is visible; e) close-up image where the transition from the darker background at the top in (d) (the lines are printed in the top-down direction) to the bright feathers is clearly distinguishable. The fast deceleration of the piezo-stage allows generation of a sharp contrast within a few hundred nanometers.

# Full-Spectrum Flexible Color Printing at the Diffraction Limit

*Patrizia Richner, Patrick Galliker, Tobias Lendenmann, Stephan J.P. Kress, David K. Kim, David J. Norris, Dimos Poulikakos*

P. Richner, Dr. P. Galliker, T. Lendenmann, Prof. D. Poulikakos
Laboratory for Thermodynamics in Emerging Technologies, ETH Zurich, Sonneggstrasse 3, 8092 Zurich, Switzerland

S. J.P. Kress, Dr. D. K. Kim, Prof. D. J. Norris
Optical Materials Engineering Laboratory, ETH Zurich, Leonhardstrasse 21, 8092 Zurich, Switzerland

**Printing: RGB-value to speed conversion**
The emission intensity of a sub-diffraction spot of printed quantum dots is a nonlinear function of the number of quantum dots, since the bottom layers may experience some quenching. Furthermore, the conversion between the 8 bit RGB value (0-255 for each color) and the spectrally measured intensity is non-linear as well. This accounts for the fact that the light sensitivity of the human eye spans several orders of magnitude and the perception of intensity is hence logarithmic.[1] The RGB value to printing speed conversion was therefore fitted with the following function:

$$v = \frac{1560 - 4.791x}{x + 11.04}$$

Where x is the RGB value (0-255) and v is the resulting velocity in $\mu$m/s, ranging from 141 $\mu$m/s to 1.3 $\mu$m/s.

**Printing: Alignment procedure**

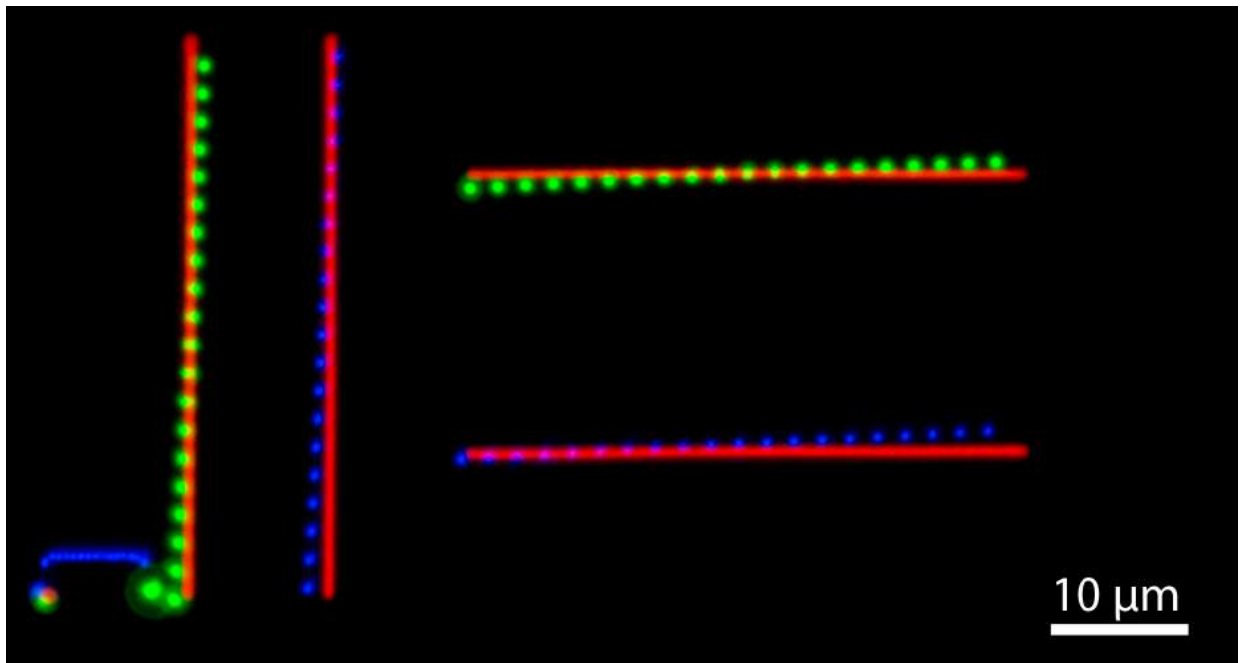

**Supporting Figure 1:** Microscope photograph showing the alignment scheme for green and blue relative to red.

A full-color, photorealistic, printed image has to be printed with three different inks and consequently with three different nozzle-printheads. The second and the third nozzle need to be aligned carefully to the first image to avoid chromatic off-sets. Since the mounted sample is left untouched on the stage, there is no rotational misalignment. Before printing the desired image, four lines (here: red) are printed at a well-defined location from the starting point, two lines for each axis. When switching to the next color, the nozzle is aligned manually to a precision of less than 2 $\mu$m. 20 points are then printed across the red line, each dot with an offset of 100 nm from its predecessor. With the help of the high-resolution iSCAT-microscopy mode[2] it is possible to determine which dot has no off-set with respect to the line. After providing this information for each axis to the printing software program, it is straight-forward to achieve an alignment error of less than 100 nm.

**CIE chromaticity diagram: Spectra to point conversion**

The points in the chromaticity diagram are calculated from the measured emission spectra. As the chromaticity diagram does by definition not contain any information about the intensity, i.e. all colors are saturated, the spectra are normalized. The conversion formulae for the conversion of a given spectrum $S(\lambda)$ are given below:

color matching functions as defined by the CIE commission:[3] $\bar{x}(\lambda)$, $\bar{y}(\lambda)$, $\bar{z}(\lambda)$

$$X = \int (S(\lambda)\bar{x}(\lambda)d\lambda$$

$$Y = \int (S(\lambda)\bar{y}(\lambda)d\lambda$$

$$Z = \int (S(\lambda)\bar{z}(\lambda)d\lambda$$

Normalization:

$$x = \frac{X}{X+Y+Z}$$

$$y = \frac{Y}{X+Y+Z}$$

$$z = \frac{Z}{X+Y+Z}$$

The z coordinate is not needed to uniquely define a point, as it can be expressed as a function of x and y: z=1-x-y.